# CCP: Conflicts Check Protocol for Bitcoin Block Security[1]

Chen Yang
*Peking University, China*
yc90011@pku.edu.cn

Haohong Wang
*TCL Research America, USA*
haohong.wang@tcl.com

**Abstract**

*In this work, we present our early stage results on a Conflicts Check Protocol (CCP) that enables preventing potential attacks on bitcoin system. Based on the observation and discovery of a common symptom that many attacks may generate, CCP refines the current bitcoin systems by proposing a novel arbitration mechanism that is capable to determine the approval or abandon of certain transactions involved in confliction. This work examines the security issue of bitcoin from a new perspective, which may extend to a larger scope of attack analysis and prevention.*

## 1. Introduction

Bitcoin [1-5] has been very popular since its invention nine years ago and so far had over 8.8 million users [3] and the market cap for Bitcoin has reached over 91B USD as of October 2017. In Bitcoin system, a peer-to-peer (P2P) network is formed via the basic element of node (or called "client" or "peer"), and each node maintains a list of other nodes' IP addresses for communicating. The operations of Bitcoin transaction (like cash payment) consistency purely rely on a mechanism called Block Chain, which is composed of batches of approved transactions that have been grouped together (called "block"). Each block contains a cryptographic hash of its predecessor, so that blocks can form (one or more) block chains. The block chain thus forms an incremental log of all transactions that have ever occurred since the first block of block chain. This way, every transfer of money can be verified by reading the log from start to end. With the growth of the block chain, newly created blocks are flooded through the network to ensure all nodes possess them, thus ideally the ownership of every fraction of a bitcoin is aware and agreed by all nodes in the system, which in reality requires tremendous efforts in maintaining information consistency and capability to handle conflictions.

Bitcoin system is designed in the rational of "the block is easy to verify but hard to create', that is, the valid blocks are required to contain a proof-of-work, which requires tremendous computation and many guessing efforts. A longest-chain rule is adopted to resolve conflicts, that is, the one requires highest aggregate difficulty of proof-of-work computations will be adopted. This way the network will eventually converge by removing conflicts.

Due to the natural of distributed system, Bitcoin system still faces many attacks [2, 6]. Double spending attack is a common one that an attacker tries to make two conflicting transactions at the same time in order to double use one single payment. On the other hand, if a miner (attacker) controls enough computational resource, he/she can mine a conflicting block chain long enough to overtake the current main chain, which is forking attack. For example, the attacker broadcast to the network a transaction in which the attacked merchant is paid by the attacker and wait the confirmation of the transaction and goods from the merchant; Then the attacker secretly mines a conflicting block chain containing a conflicting transaction which pays the attacker himself instead of the merchant; After that the attacker extends the secret block chain until it overtakes the public chain. As another example, (which has not been discussed before) an attacker collects transactions that are spent but not mined in block chain, and then mines a conflicting block containing conflicting transactions, which pay to the attacker instead of their original receivers. Clearly, this attack is much severer because the attacker changes transactions that are sent by anyone.

In this work, we propose a novel mechanism, called Conflicts Check Protocol (CCP) that can reduce the chance of many attacks to minimum. CCP is based on an observation that many attacks are having a common symptom that there are more than one transaction sharing a same input, thus it is critical to effectively detect such risky symptom. CCP proposes a novel arbitration process to solve these conflicts after the detection. The paper is organized as follows: in the second section, the common symptom of many attacks is demonstrated. In section 3, the CCP mechanism is proposed. The future value, opening topics, and extensibility of CCP are discussed in the last section with the final conclusion.

## 2. Conflicts Symptom of Many Attacks

As mentioned earlier, a transaction containing bitcoin needs to be validated by all nodes of the system. A transaction normally contains one or more inputs that represent the senders of the bitcoins, and one or more

---

outputs that represent the receivers of the bitcoins. A received bitcoin cannot be used until the corresponding transaction has been validated by other nodes. The transactions are flooded within Bitcoin P2P network so that each node gets notified for these transactions. All nodes need to repeat verification work of transaction independently, including checking double spending and other validation rules. Like other attacks, double spending happens when conflicting transactions attempt to transfer the exact same bitcoin to different destinations. When there are conflicting transactions (i.e., they carry the same bitcoin) in a P2P network, the node observing conflicts typically approve the first transaction it receives based on flood algorithm, as it does not have a mechanism to report the confliction observed after receiving the second transaction with a same input. Block chain can solve some attacks having this symptom (i.e., conflicts) but not all of them, for example, forking attacks cannot be solved completely. This important symptom can be caught by CCP to fix these attacks.

Clearly, there are other ways attackers choose to use [7-8] that may not have such symptom. In this work, we are focus on the type of symptom that can be easily detected by each node based on the transaction conflicts observed to have at least one same input.

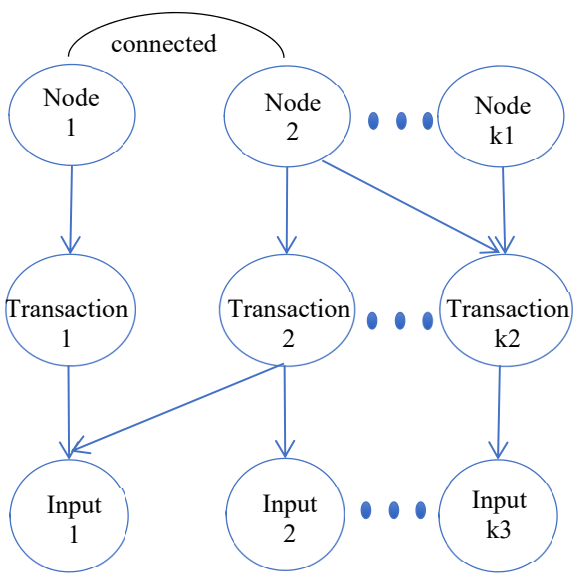

Figure 1. Example showing relationship among nodes, transactions and inputs in Bitcoin system

As shown in Fig. 1, there are multiple nodes in the system (e.g., *k1* nodes in total), each node approves multiple transactions (assuming in total of *k2* transactions in the system), which is indicated by the arrow in the figure. It is also important to realize that Transaction 1 and Transaction 2 in the figure are conflicting with each other as they share a same input (i.e., Input 1). Lines indicate the relationship of connect. Node 1 is connected to Node 2 (and some other nodes). Node 2 is also connected to Node 1.In the next section, we will discuss the mechanism used to remove chances of attacks after detecting this type of symptom related to transaction conflicts.

## 3. CCP – Conflicts Check Protocol

CCP is built on top of Bitcoin system, it modifies the system in the following aspects:

- The interactions between node and transaction: in Bitcoin system, a transaction is either approved or refused by a node; in CCP, the approval state is split into 2 states, namely, tentatively approved and final approved. A node can tentatively approve a transaction, but later on, if conflict is detected, the node will report the observed conflict to an arbitrator, where an arbitration mechanism will communicate with connected nodes and help the node to decide whether continue tentatively approving the transaction or tentatively refuse it.

- In CCP, conceptually the time domain is equally divided into many intervals, and there is an action point in each time interval that all nodes can take actions only at these points, thus at every action point, nodes of the whole network conduct an update (e.g., tentatively approve a transaction, raise arbitration, or receive the arbitration outcome from other connected nodes at last action point and change its own arbitration). For example, when a node detects a transaction conflict at an input, it triggers the arbitration mechanism, when the arbitration outcome is returned, it may tentatively approve the later-coming transaction at the input, or tentatively approve the earlier-tentatively approved one, or decide to abandon all transactions at the input. It is important to realize that in any time, it is not possible for a node to tentatively approve more than one transaction for a specific input.

- An arbitrator and corresponding arbitration mechanism is deployed to resolve conflicts. As shown in Fig. 2, an example of arbitrator for Input 1 of Node 1 is demonstrated, where Node 1 realized that both Transaction 1 and Transaction 2 it received contains the same Input 1 and thus raised a confliction. All transactions that have input 1 (i.e., Transaction 1 and Transaction 2), are treated as transactions in confliction by the arbitrator. The arbitrator uses a few rounds of voting to determine whether there is a transaction can be approved at this input or all transactions need to be abandoned. There are only two possible outcomes of arbitration in each round: either a transaction becomes the one in Domination (see example in Fig. 2 that Transaction 1 is judged to win out other transactions) and then the node tentatively approves that transaction, or no one in domination and thus all transactions in conflict are abandoned. When starting the arbitration mechanism of input 1, node 1 tentatively approves the transaction it receives first that contains input 1. At each action point mentioned above, the arbitrator communicates with the connected nodes involved (as shown in Fig. 2 the node 2 group and the node 3 group), and determine the arbitration outcome. After several rounds, at each input, each node has big enough possibility to have the same arbitration (domination or abandon) as each other node's (see below the mathematically proof), then the tentative approval (and tentative abandon) becomes final approval (final abandon) and cannot change any more. Only if the node finally approves a transaction at every input that is contained in this transaction, the node finally approves this transaction; the node will finally refuse the transaction in any other final arbitration outcome.

✦ In CCP, only a fixed number of connected nodes (say $H$) are considered in the voting process of arbitration. Although the node in classic Bitcoin system is allowed to connect with 8-128 nodes, our experiments indicate that the selection of $H$ does not have significant impact on our conclusion. Once the $H$ is determined by the system, it will be utilized in all arbitration processes.

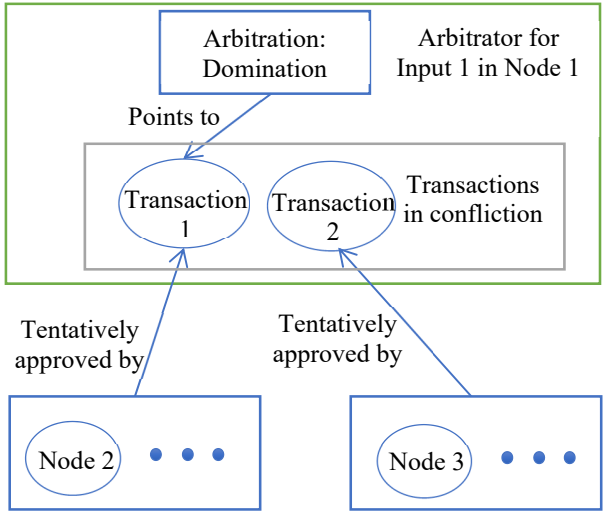

Figure 2. An example of the arbitration mechanism

It is important to realize that the arbitration made by the arbitrator according to the voting process by the connected nodes with their tentatively approvals at an input is theoretically with high confidence level. Let us denote by $m$ the number of transactions in conflict currently handling at the arbitrator of an input, $h[1]$, $h[2]$ …$h[m]$ the numbers of connected nodes that tentatively approved these $m$ different transactions at that input respectively. According to flood algorithm, each node has to tentatively approve a transaction or choose to abandon at the input at each action point. Let us also denote by $h[m+1]$ the number of connected nodes that choose abandon. $H$ is the sum of all connected nodes and then the arbitration process can be formally defined by using the following pseudocodes :

0   Initialize – start timer (from 0)
1   while (timer++ < Action Point E) do
2       outcome = "abandon"
3       update h[*] based on information periodically received from connected nodes during last time interval
4       for i=1 to m do
5           if (h[i]/H > r)   // r is a threshold
6               then outcome = {"domination", i}; break;

In this work, we assume the P2P network is large enough and thus the connections between nodes and the behavior model of nodes are quite random. In a random process, we can assume that any 2 randomly picked nodes have similar probability to tentatively approve a single transaction. Let us denote by $p$ the probability of each connected node to tentatively approve a transaction at an input, then $h[i]$ obeys Binomial Distributions [9], which has the following probability:

$$P(h[i]=x)=b(x,H,p)= C_H^x *p^x *(1-p)^{H-x} \quad (1)$$

Assuming $p_1$ is the possibility that a node tentatively approves this transaction in arbitration process after a time interval, then

$$p_1=P(\tfrac{h[i]}{H}>r) \quad (2)$$

It is important to realize that $p_1$ is the value of $p$ after a time period. According to central limit theorem, Binomial Distribution can be approximated to Normal distribution. So

$$p_1=P(\tfrac{h[i]}{H}>r)\rightarrow P(u>\tfrac{Hr-Hp}{\sqrt{Hp(1-p)}}) |u\sim N(0,1)) \quad (3)$$

that is,

$$p_1= \tfrac{1}{\sqrt{2\pi}} \int_{\tfrac{Hr-Hp}{\sqrt{Hp(1-p)}}}^{\infty} exp(-\tfrac{u^2}{2}) du \quad (4)$$

If we use a function called $P_1()$ to represent the relationship between $p_1$ and original p in Eq. (4), like $p_1 = P_1(p)$, it can be observed that $P_1(p)$ is monotone, and $P_1(p)$ is the also the value of $p$ after 1 time interval in arbitration process. In the same manner, $P_1(P_1(p))$ is the value of $p$ after 2 time intervals in arbitration process; $P_1(P_1(P_1(p)))$ is the value of $p$ after 3 time intervals, and $P_1^n(p)$ is the value $p$ after $n$ time intervals. Following equation (4), we are also able to derive the exact function of $P_1^n(p)$. When we compare the functions of $P_1^n(p)$ in Fig. 3 with different value of n, we observe that the slope of function gets steeper with the increase of $n$.

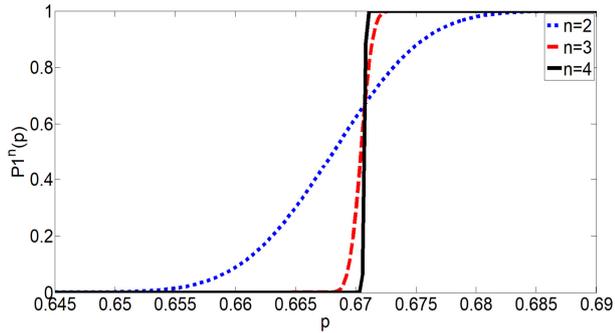

Figure 3. The relationship of $p$ and $P_1^n(p)$ (n=2 3 and 4) when $H$=100

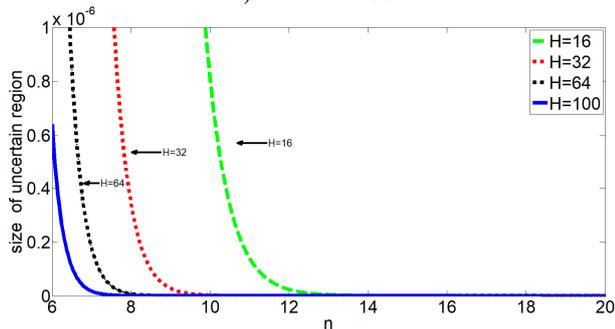

Figure 4. The relationship of $n$ and the length of uncertain region with different voting number

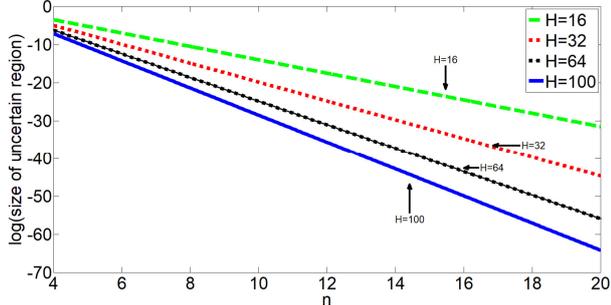

Figure 5. The relationship of $n$ and log(the length of uncertain region) with different voting number

In Fig. 3, when $p$>0.69, $P_1(P_1(p))$=1, which means all nodes tentatively approve the transaction after 2 time intervals and that won't change in the rest of the time. When $p$<0.64, $P_1(P_1(p))$ =0, which means all nodes tentatively do not approve the transaction after 2 time intervals and that does not change in the rest of the time. When 0.64<$p$<0.69, nodes may not get common view during arbitration process in 2 time intervals. So (0.64,0.69) is the uncertain region of arbitration process after 2 time intervals. It is important to realize that the size of uncertain region is getting smaller and smaller with the increase of $n$, it becomes 0.005 when $n$=3, 0.0008 when n=4, and smaller than $10^{-6}$ when n>=6 (as shown in Fig. 4), and the size decrease very fast when $n$ further increase (as shown in Fig. 5). In Figs. 4 and 5, the relationship of $n$ and $H$ are demonstrated. Clearly, smaller $H$ (e.g. $H$=16) corresponds to longer convergence speed, however, all settings of $H$ indicates that the size of uncertain region can be reduce to a value very close to 0 as the increase of $n$.

It is not hard to understand that when the size of uncertain region is too small to become negligible (e.g., when $n$>12), we can claim that at certain action point (after *several* rounds of voting) CCP has full confidence that all nodes will be consistent on either select a single transaction for approval, or choose to abandon all transactions in confliction.

## 4. Conclusion and Discussions

In this paper, a novel approach to detect and resolve attacks, called CCP, is proposed. It is based on an observation that in many attacks a common symptom appears, that is, there is more than one transaction sharing a same input. After detecting this symptom, a novel arbitration mechanism is proposed in CCP to resolve the conflicts.

This work is still in its very early stage, some of its assumption has not been fully verified in simulation, for example, the model of using a same probability for each connected node to tentatively approve a transaction at an input. However, it is worth noting that CCP opens a new door for bitcoin system, and we believe it will play significant roles in maintaining consistency, and it may also be used for other attacks detection and resolution by covering more extensive symptoms discovery.

# Appendix
# Another kind of forking attack—may be 1 million severer than classic forking attack

Another kind of forking attack, which can be called custom block attack, has not been noticed before: an attacker collects transactions that are spent but not mined in block chain, and then mines a conflicting block containing conflicting transactions, which pay to the attacker instead of their original receivers. In classic forking attack, the attacker only changes transaction that is sent by himself while in custom block attack, the attacker changes transactions that are sent by anyone. Both two kinds of forking attack are changing the transactions' outputs in blocks so custom block attack is operational. The results discussing classic forking attack in [6] can be applied to custom block attack easily but custom block attack is much severer than double spending attack with two reasons: 1, a block usually has over a thousand transactions so a succeed custom block can make more than a thousand times of profits than double spending attack; 2, custom block attack may be succeeded with fewer secret blocks mined. Generally double spending in forking attack needs the attacker mines a secret block chain with at least 6 blocks longer than the public chain [5] [10], but the length requirement of the secret block chain in custom block attack is decided by miners instead of the merchant that gets paid. In default situation, miners should accept the longest block chain so custom block attack may succeed with only 1 secret block faster than public chain, corresponding 1-confirmation attack in [6]. The possibility of 1-confirmation attack (custom block attack) is about 1 thousand times higher than 6-confirmation attack (double spending) so the benefit cost ratio of custom block attack may be 1 million times higher than double spending in forking attack.